\documentclass[letterpaper,twocolumn,10pt]{article}
\usepackage{usenix}

\usepackage{pifont}
\usepackage{nicefrac}
\usepackage{siunitx}
\usepackage{array,framed}
\usepackage{booktabs}
\usepackage{
  float,
  epsfig,
  wrapfig,
  graphics,
  graphicx,
  subcaption
}
\usepackage{textcomp}
\usepackage{amssymb}
\usepackage{setspace}
\usepackage{latexsym}
\usepackage{enumerate}
\usepackage{enumitem}
\usepackage{algorithm}
\usepackage{algpseudocode}
\usepackage{listings}
\usepackage{graphics}
\usepackage{xparse} 
\usepackage{xspace}
\usepackage{multirow}
\usepackage{csvsimple}
\usepackage{balance}
\usepackage[many]{tcolorbox}

\usepackage{
  tikz,
  pgfplots,
  pgfplotstable
}

\usetikzlibrary{
  shapes.geometric,
  arrows,
  external,
  pgfplots.groupplots,
  matrix
}

\pgfplotsset{compat=1.9}

\usepackage{mathtools,}

\DeclareMathAlphabet{\mathcal}{OMS}{cmsy}{m}{n}

\DeclareGraphicsExtensions{%
    .png,.PNG,%
    .pdf,.PDF,%
    .jpg,.mps,.jpeg,.jbig2,.jb2,.JPG,.JPEG,.JBIG2,.JB2}

\usepackage{xparse}
\newcommand{\bnm}{\begin{newmath}}
\newcommand{\enm}{\end{newmath}}

\newcommand{\bea}{\begin{eqnarray*}}%
\newcommand{\eea}{\end{eqnarray*}}%

\newcommand{\bne}{\begin{newequation}}
\newcommand{\ene}{\end{newequation}}

\newcommand{\bal}{\begin{newalign}}
\newcommand{\eal}{\end{newalign}}

\newenvironment{newalign}{\begin{align}%
\setlength{\abovedisplayskip}{4pt}%
\setlength{\belowdisplayskip}{4pt}%
\setlength{\abovedisplayshortskip}{6pt}%
\setlength{\belowdisplayshortskip}{6pt} }{\end{align}}

\newenvironment{newmath}{\begin{displaymath}%
\setlength{\abovedisplayskip}{4pt}%
\setlength{\belowdisplayskip}{4pt}%
\setlength{\abovedisplayshortskip}{6pt}%
\setlength{\belowdisplayshortskip}{6pt} }{\end{displaymath}}

\newenvironment{newequation}{\begin{equation}%
\setlength{\abovedisplayskip}{4pt}%
\setlength{\belowdisplayskip}{4pt}%
\setlength{\abovedisplayshortskip}{6pt}%
\setlength{\belowdisplayshortskip}{6pt} }{\end{equation}}

\newcounter{ctr}

\newcounter{mytable}
\def\mytable{\begin{centering}\refstepcounter{mytable}}
\def\endmytable{\end{centering}}

\newcounter{myfig}
\def\myfig{\begin{centering}\refstepcounter{myfig}}
\def\endmyfig{\end{centering}}

\newlength{\saveparindent}
\newlength{\saveparskip}
\newcommand{\E}{{\rm I\kern-.3em E}}

\newcommand{\appref}[1]{\mbox{Appendix~\ref{#1}}}

\renewcommand{\eqref}[1]{\mbox{Equation~(\ref{#1})}}

\def \part {part}

\def \blackslug{\hbox{\hskip 1pt \vrule width 4pt height 8pt
    depth 1.5pt \hskip 1pt}}
\def \qed{\quad\blackslug\lower 8.5pt\null\par}

\newcounter{mynote}[section]

\newcommand\ignore[1]{}

\newcounter{rcnote}[section]

\newcounter{mrnote}[section]

\newcounter{fknote}[section]

\newcounter{anote}[section]

\DeclareMathSymbol{\mlq}{\mathord}{operators}{``}
\DeclareMathSymbol{\mrq}{\mathord}{operators}{`'}

\newcommand{\rhf}[2]{R_{f, \gamma}}

\DeclareDocumentCommand{\edist}{o o}{
  \ensuremath{
    \IfNoValueTF{#1}{{d}}{{\sf d}(#1,#2)}
  }
}

\newcommand{\olrk}[1]{\ifx\nursymbol#1\else\!\!\mskip4.5mu plus 0.5mu\left(\mskip0.5mu plus0.5mu #1\mskip1.5mu plus0.5mu \right)\fi}

\newcommand{\tool}{\textsc{PentestGPT v2}\xspace}

\NewDocumentCommand{\indseq}{ O{1} O{r} }{{#1}\ldots {#2}}

\begin{document}

\def\thetitle{What Makes a Good LLM Agent for Real-world Penetration Testing?}
\title{\thetitle}

\newcommand{\mkntu}[0]{{{$^1$}}}
\newcommand{\mkunsw}[0]{{{$^2$}}}
\newcommand{\mksmu}[0]{{{$^3$}}}
\newcommand{\mkcfar}[0]{{{$^4$}}}
\newcommand{\mkthu}[0]{{{$^5$}}}

\author{
{\rm Gelei Deng}\mkntu \rm,
{\rm Yi Liu}\mkntu \rm,
{\rm Yuekang Li}\mkunsw\rm,
{\rm Ruozhao Yang}\mksmu \rm, 
{\rm Xiaofei Xie}\mksmu \rm, \\
{\rm Jie Zhang}\mkcfar \rm, 
{\rm Han Qiu}\mkthu \rm,
{\rm Tianwei Zhang}\mkntu \rm \\
\mkntu {Nanyang Technological University},
\mkunsw {University of New South Wales},
\mksmu {Singapore Management University}, \\
\mkcfar {CFAR, A*STAR, Singapore},
\mkthu {Tsinghua University}\medskip
}


\maketitle

\begin{abstract}
LLM-based agents show promise for automating penetration testing, yet the reported performance varies widely across systems and benchmarks.
We analyze 28 LLM-based penetration testing systems and evaluate five representative implementations across three benchmarks of increasing complexity.
Our analysis reveals two distinct failure modes: \emph{Type A failures} stem from capability gaps (missing tools, inadequate prompts) that engineering readily addresses, while \emph{Type B failures} persist regardless of tooling due to planning and state management limitations.
We show that Type B failures share a root cause that is largely invariant to the underlying LLM: agents lack real-time task difficulty estimation. 
As a result, agents misallocate effort, over-commit to low-value branches, and exhaust context before completing attack chains.

Based on this insight, we present \tool, a penetration testing agent that couples strong tooling with difficulty-aware planning.
A Tool and Skill Layer eliminates Type A failures through typed interfaces and retrieval-augmented knowledge.
A Task Difficulty Assessment (TDA) mechanism addresses Type B failures by estimating tractability through four measurable dimensions (horizon estimation, evidence confidence, context load, and historical success) and uses these estimates to guide exploration-exploitation decisions within an Evidence-Guided Attack Tree Search (EGATS) framework.
\tool achieves up to 91\% task completion on CTF benchmarks with frontier models (39 to 49\% relative improvement over baselines) and compromises 4 of 5 hosts on the GOAD Active Directory environment versus 2 by prior systems.
These results show that difficulty-aware planning yields consistent end-to-end gains across models and addresses a limitation that model scaling alone does not eliminate.
\end{abstract}

\section{Introduction}
\label{sec:intro}

Penetration testing is essential for assessing organizational security, yet the demand for skilled practitioners far exceeds supply. The ISC2 Cybersecurity Workforce Study estimates a global shortfall of 4.7 million cybersecurity professionals~\cite{isc2workforce2024}. This gap, together with the labor-intensive nature of manual testing, has driven interest in large language model (LLM)–based automation.

Recent systems report strong results on benchmarks such as Capture-the-Flag challenges and Hack The Box (HTB) environments~\cite{deng2024pentestgpt,chen2024pentestagent,autopt2024,vulnbot2025,xoffense2024}, and emerging work has demonstrated real-world impact, including the discovery of exploitable vulnerabilities in production software~\cite{bigsleep2024,heelan2025smb}. However, reported task completion rates range from single digits under naive prompting to 40--80\% with more sophisticated architectures~\cite{termibench2024,autopenbench2024}, raising a central question: \emph{what drives these performance differences, and what limitations remain?}

To answer this question, we conduct a systematic analysis of 28 LLM-based penetration testing systems and evaluate five representative solutions across three benchmarks of increasing complexity.
Our analysis yields two findings. First, existing systems are optimized to address the limitations of specific LLMs. For example, context summarization and RAG-augmented tooling are designed to compensate for transient LLM constraints of limited context windows and poor tool knowledge. Benefits brought by these designs quickly diminish as models improve: performance gaps across solutions compress by over half when backbone models upgrade from GPT-4o to GPT-5.
Second, failures partition into two categories: \emph{Type A failures} (capability gaps) stem from missing tools and knowledge addressable through engineering, while \emph{Type B failures} (complexity barriers) persist regardless of tooling due to planning and state management limitations. Existing systems predominantly target Type A failures, achieving strong results on simple tasks but failing on multi-step scenarios where Type B failures dominate. This indicates that the architectures of existing penetration testing systems are not designed to complement the improvements of LLMs. Their contributions erode as models advance, rather than compounding with improved capabilities.

We trace Type B failures to a missing capability: \textit{existing penetration testing agent designs cannot assess task difficulty in real time}.
This manifests in several ways: agents commit prematurely to unproductive branches because they cannot estimate whether a path requires 3 or 30 steps; they fail to transition from reconnaissance to exploitation because they lack metrics for evidence sufficiency; they experience context forgetting because they do not monitor context consumption.
Human pentesters handle these problems through intuition built from experience. LLM agents lack equivalent mechanisms for difficulty-aware decision making.
We validate this diagnosis through controlled evaluation: augmenting agents with difficulty assessment reduces the Type B failure rate from 58\% to 27\% while Type A rate remains unchanged, confirming that this enhancement addresses the root cause. 

We present \tool, designed around these two findings.
To eliminate Type A failures, an extensible \emph{Tool and Skill Layer} provides typed interfaces for 38 security tools and skill compositions that encode expert attack patterns. To address Type B failures, we introduce penetration testing \emph{Task Difficulty Assessment} (TDA), a mechanism that estimates task tractability through four measurable dimensions: horizon estimation, evidence confidence, context load, and historical success rate.
TDA is integrated into an \emph{Evidence-Guided Attack Tree Search} algorithm that guides exploration-exploitation decisions and prunes branches when paths become intractable.
With these mechanisms, \tool dynamically pivots between attack paths based on real-time difficulty signals. It abandons unproductive branches before they exhaust the context budget and commits to exploitation only when evidence confidence justifies the investment.
A retrieval-augmented \emph{Memory Subsystem} maintains structured state external to the LLM context, which prevents the context forgetting that derails extended attack campaigns.

We evaluate \tool across three benchmarks at different levels of realism, from CTF challenges to enterprise Active Directory environments.
On XBOW~\cite{xbow2024} (104 web security tasks), \tool achieves 91\% peak task completion (89\% mean) with Claude Opus 4.5, a 49\% relative improvement over the best baseline (61\%).
On the PentestGPT~\cite{deng2024pentestgpt} Benchmark (13 HTB/VulnHub machines), \tool roots 12 of 13 machines, solving Hard-rated targets where baselines become stuck at initial steps.
On GOAD (5-host Active Directory environment), \tool compromises 4 of 5 hosts compared to at most 2 for prior systems, with successful lateral movement and credential chaining across domain boundaries.
Ablation studies confirm that each component contributes distinctly: the Tool Layer dominates on short-horizon tasks, while TDA-EGATS and Memory provide the gains on multi-step scenarios.

Despite these results, hard challenges remain.
Our evaluation shows that novel exploitation requiring creative reasoning, adversarial environments with deceptive defenses, and extended multi-week campaigns exceed current LLM capabilities.
These limitations suggest that fully autonomous penetration testing remains distant.
We discuss these boundaries and propose evaluation methodologies that distinguish tractable from intractable challenges, so that the community can focus effort where architectural innovation is most likely to help.

In summary, we make the following contributions:

\begin{itemize}[leftmargin=*,noitemsep,topsep=5pt,parsep=0pt,partopsep=0pt]
\item \textbf{Systematic analysis of LLM agent failures} (\S\ref{sec:study}). We analyze 28 systems and evaluate five implementations across three benchmarks, showing that existing architectures optimize for transient model constraints rather than persistent task challenges, and identifying two failure categories (Type A capability gaps and Type B complexity barriers) whose root causes require distinct solutions.

\item \textbf{\tool} (\S\ref{sec:methodology}). We present a system addressing both failure types: a Tool and Skill Layer for Type A failures, and Task Difficulty Assessment integrated into Evidence-Guided Attack Tree Search for Type B failures.

\item \textbf{Evaluation across three benchmarks} (\S\ref{sec:evaluation}). \tool achieves 91\% on CTF benchmarks (49\% improvement), roots 12/13 machines on realistic targets, and compromises 4/5 hosts on enterprise AD, doubling baseline performance.

\item \textbf{Design principles} (\S\ref{sec:discussion}). We analyze remaining barriers (novel exploitation, adversarial robustness) and propose evaluation methodologies that separately assess Type A and Type B performance.

\item \textbf{Open-source artifacts}. We release \tool's implementation, tool interfaces, and evaluation scripts to support reproducibility~\cite{excalibur2025artifact}.
\end{itemize}


\section{Background}
\label{sec:relwork}

\subsection{Penetration Testing}
\label{sec:bg-pentest}

Penetration testing identifies security vulnerabilities by simulating real-world attackers in blackbox/greybox scenarios.
Standard methodologies decompose engagements into phases: \emph{reconnaissance} (information gathering), \emph{enumeration} (identifying services and entry points), \emph{exploitation} (gaining access), and \emph{post-exploitation} (privilege escalation and lateral movement)~\cite{ptes2012,owasp2021}.
This workflow follows a characteristic search pattern: \emph{breadth-first exploration} over attack surfaces followed by \emph{depth-first exploitation} along promising paths.
Testers continuously decide which paths to pursue, when to abandon unproductive avenues, and how to integrate new discoveries.
This interleaving of exploration and exploitation motivates our design (\S\ref{sec:methodology}).

\subsection{Benchmarking Penetration Testing}

Evaluating penetration testing capabilities presents methodological challenges.
Real-world engagements involve social engineering, multi-target reconnaissance, and complex business logic that cannot be easily replicated, while commercial tests produce confidential reports tied to proprietary systems.
Standardized benchmarks address these constraints: \emph{VulnHub}~\cite{vulnhub} provides downloadable vulnerable VMs, \emph{HTB}~\cite{hackthebox} offers curated machines spanning difficulty levels, and \emph{CTF} competitions present challenges across web exploitation, cryptography, and binary exploitation.

Benchmarks differ from real-world targets in important ways.
CTF challenges are designed to be solvable with a single attack path, whereas real systems may have no exploitable vulnerabilities or require broad discovery across a large attack surface.
GOAD (Game of Active Directory)~\cite{goad2024} is the closest approximation to realistic enterprise environments among current benchmarks, requiring chained attack techniques across multi-domain Windows networks, though it still abstracts away social engineering and time pressure.
We interpret benchmark results as measuring specific technical capabilities rather than predicting overall real-world effectiveness.

\subsection{LLM-Based Agents}
\label{sec:bg-llm-agents}

The standard approach for deploying LLMs as autonomous agents augments them with \emph{tool use}~\cite{wolflein-etal-2025-llm} that invokes external functions such as shell commands or APIs, and \emph{agentic scaffolding} that structures the interaction loop~\cite{jimenez2024swebench,zhou2024webarena}.
Penetration testing is a natural application domain for such agents: it requires combining extensive domain knowledge with sequential decision-making, tool orchestration, and adaptive strategy.
Early work explores LLMs as copilots suggesting next steps to human operators~\cite{deng2024pentestgpt,shao2024ctfllm}, whereas more recent systems position LLMs as autonomous agents executing reconnaissance, exploitation, and post-exploitation workflows~\cite{autopt2024,chen2024pentestagent,vulnbot2025}.
These agents must handle heterogeneous tool outputs, maintain coherent strategies across many interaction steps, and decide when to pivot between attack paths.
These challenges push against the limits of current LLM capabilities.
Similar limitations appear in software engineering~\cite{jimenez2024swebench} and web navigation~\cite{zhou2024webarena}, suggesting that the barriers are not specific to penetration testing.

\section{Understanding LLM Agent Failures}
\label{sec:study}

\emph{How far are we from achieving real-world penetration testing with LLM agents?}
To answer this question, we conduct an empirical analysis of existing LLM-based penetration testing systems.
Our goals are to (1) understand what drives reported performance improvements, (2) identify failure modes through controlled evaluation, and (3) establish a framework for distinguishing tractable tasks from intractable challenges.

\subsection{Taxonomy and Evaluation of LLM-based Penetration Testing}
\label{sec:study-survey}

We survey LLM-based penetration testing systems, identifying 28 candidates published between 2023--2025.
Inclusion criteria require systems to use LLMs as a core component and target penetration testing or CTF challenges; we exclude vulnerability detection without exploitation and commercial systems without published details.
Of 28 candidates, 10 meet our criteria, with the list in Appendix~\ref{sec:appendix-systems}.

\subsubsection{Taxonomy}

We summarize each system along four dimensions: \emph{architecture} (multi-agent, human-in-the-loop), \emph{tool integration} (function calls, MCP~\cite{mcp2024}), \emph{knowledge sources} (Retrieval-Augmented Generation (RAG), fine-tuned), and \emph{planning} (reactive, task trees, state machines, memory trees).
Table~\ref{tab:taxonomy} summarizes representative systems across three architectural families: human-in-the-loop copilots like PentestGPT~\cite{deng2024pentestgpt}, single-agent systems like AutoPT~\cite{autopt2024}, and multi-agent systems like PentestAgent~\cite{chen2024pentestagent}, VulnBot~\cite{vulnbot2025}, and Cochise~\cite{happe2025cochise}.

\begin{table}[t]
\centering
\caption{Taxonomy of LLM-based penetration testing systems 
}
\label{tab:taxonomy}
\resizebox{\linewidth}{!}{
\begin{tabular}{llllll}
\toprule
\textbf{System} & \textbf{Year} & \textbf{Arch.} & \textbf{Tools} & \textbf{Know.} & \textbf{Planning} \\
\midrule
PentestGPT~\cite{deng2024pentestgpt} & 2024 & Workflow & Shell & Prompt & Task tree \\
AutoPT~\cite{autopt2024} & 2024 & Single & Shell & Prompt & State mach. \\
RapidPen~\cite{rapidpen2025} & 2025 & Single & Shell & RAG & ReAct \\
PentestAgent~\cite{chen2024pentestagent} & 2024 & Multi & Func. & RAG & Phase \\
VulnBot~\cite{vulnbot2025} & 2025 & Multi & Shell & Prompt & Tri-phase \\
xOffense~\cite{xoffense2024} & 2024 & Multi & Shell & Fine-tune & Multi-phase \\
TermiAgent~\cite{termibench2024} & 2024 & Multi & Shell & RAG & Mem. tree \\
Cochise~\cite{happe2025cochise} & 2025 & Multi & Shell & Prompt & Hierarchical \\
\bottomrule
\end{tabular}}
\end{table}

\subsubsection{Evaluation Setup}

We evaluate five representative open-source systems:
\textbf{PentestGPT}~\cite{deng2024pentestgpt} (copilot),
\textbf{AutoPT}~\cite{autopt2024} (single-agent),
\textbf{PentestAgent}~\cite{chen2024pentestagent} (multi-agent with RAG),
\textbf{VulnBot}~\cite{vulnbot2025} (multi-agent tri-phase), and
\textbf{Cochise}~\cite{happe2025cochise} (Active Directory (AD)-focused).
Benchmarks span three realism levels:
\textbf{XBOW}~\cite{xbow2024} (104 web challenges: SQL injection (SQLi), cross-site scripting (XSS), auth bypass),
\textbf{PentestGPT Benchmark}~\cite{deng2024pentestgpt} (13 machines from HTB and VulnHub requiring end-to-end penetration testing), and
\textbf{GOAD}~\cite{goad2024} (5-host multi-domain AD requiring chained attacks).

For each system-benchmark pair, we evaluate with GPT-4o, GPT-5, Gemini-3-Flash, and Claude Sonnet 4 to assess model vs.\ architecture contributions.
We include GPT-4o (the model generation most existing systems were optimized for) alongside newer models to examine how architectural advantages evolve as underlying capabilities improve.
\S\ref{sec:evaluation} evaluates \tool{} with a different model set (GPT-5.2, Opus 4.5, Gemini 3 Pro) selected specifically for thinking mode support, enabling controlled comparison of extended reasoning.
We set temperature to zero and report best-of-three trials following prior work~\cite{deng2024pentestgpt,autopenbench2024}, since penetration testing is inherently non-deterministic. 

\subsection{Findings}
\label{sec:study-findings}

Table~\ref{tab:eval-results} summarizes task completion rates across all system-model-benchmark combinations. We provide in-depth experimental results analysis below.

\begin{table*}[t]
\centering
\caption{Task completion rates across systems, models, and benchmarks. XBOW: task completion (\%); PentestGPT Benchmark: machines rooted (/13); GOAD: hosts compromised (/5).}
\label{tab:eval-results}
\small
\begin{tabular}{l|cccc|cccc|cccc}
\toprule
& \multicolumn{4}{c|}{\textbf{XBOW (104 tasks)}} & \multicolumn{4}{c|}{\textbf{PentestGPT-Ben (13 machines)}} & \multicolumn{4}{c}{\textbf{GOAD (5 targets)}} \\
\textbf{System} & GPT-4o & GPT-5 & Gem. & Claude & GPT-4o & GPT-5 & Gem. & Claude & GPT-4o & GPT-5 & Gem. & Claude \\
\midrule
PentestGPT & 27 & 42 & 36 & 39 & 5 & 7 & 6 & 6 & 0 & 1 & 1 & 1 \\
AutoPT & 28 & 40 & 35 & 37 & 4 & 7 & 6 & 6 & 0 & 1 & 0 & 0 \\
PentestAgent & 34 & \textbf{49} & 42 & \textbf{46} & \textbf{6} & 7 & 6 & 6 & 0 & 1 & 0 & 1 \\
VulnBot & \textbf{39} & 45 & \textbf{44} & \textbf{46} & \textbf{6} & \textbf{8} & 6 & \textbf{7} & 0 & 1 & 0 & 1 \\
Cochise & 34 & 43 & 39 & 39 & 4 & 4 & 4 & 4 & \textbf{1} & \textbf{2} & \textbf{2} & \textbf{2} \\
\bottomrule
\end{tabular}
\end{table*}

\subsubsection{Agent Architecture Convergence}
Despite two years of agent design innovation, performance differences between systems compress with state-of-the-art models.
On XBOW with GPT-4o, completion rates range from 27\% to 39\% across five systems, a 44\% relative spread that reflects meaningful architectural distinctions.
With GPT-5, this gap narrows to 22.5\% (40--49\%); similar convergence appears on the PentestGPT Benchmark, where the spread shrinks from 2 points with GPT-4o (4--6 machines) to 1 point with GPT-5 (7--8 machines).

This convergence points to a limitation in how existing agents were designed: they address \emph{transient} model constraints rather than \emph{persistent} task challenges.
Consider the techniques these systems employ.
PentestGPT's summarization module compensates for limited context windows, a constraint that largely dissolves as models gain native million-token support.
Multi-agent architectures with role separation (e.g., reconnaissance agent, exploitation agent) work around weak instruction-following, yet frontier models handle complex multi-step prompts without explicit decomposition.
RAG pipelines for tool documentation address poor parametric knowledge of security tools, yet recent models have much stronger baseline knowledge of common exploitation techniques and penetration testing tools.
These ``innovations'' are workarounds for 2023-era model limitations, not solutions to persistent penetration testing challenges.

What distinguishes transient from persistent challenges?
Transient challenges diminish as models improve: context capacity, instruction adherence, tool-use reliability, and domain knowledge all scale with model capability.
Persistent challenges, by contrast, remain regardless of raw model power: long-horizon planning across 10+ exploitation steps, principled exploration-exploitation decisions, maintaining state external to degrading context, and real-time assessment of task difficulty.
These challenges arise from the \emph{structure of penetration testing tasks}, not from model limitations, and thus require architectural solutions that \emph{complement} rather than \emph{compensate for} underlying models.

The Cochise case shows this distinction from a different angle.
Cochise's AD-specific attack primitives (Kerberoasting, NTLM relay, BloodHound integration) are capability additions that models cannot replicate through improved reasoning alone.
However, this specialization comes at the cost of generality: Cochise underperforms on XBOW and the PentestGPT Benchmark (34\% and 4/13 with GPT-4o) compared to general-purpose systems like VulnBot (39\% and 6/13), while leading on GOAD by leveraging domain-specific knowledge unavailable to other systems.
Neither approach, compensating for model limitations nor adding domain-specific capabilities, addresses the persistent challenge of navigating complex attack graphs.

\begin{tcolorbox}[left=1mm, right=1mm, top=0.5mm, bottom=0.5mm, arc=1mm]
\textbf{Finding 1:} Existing penetration testing agents address transient model limitations rather than persistent task challenges. As models evolves, benefits brought by architectural distinctions compress. Durable agent value should address challenges that persist across model evolution.
\end{tcolorbox}

\subsubsection{Two Distinct Failure Categories}
To understand \emph{why} systems fail rather than merely \emph{how often}, we analyze 200 execution traces from unsuccessful attempts (40 per system), sampling proportionally across benchmarks.
Two researchers independently coded failure modes using open coding, then reconciled disagreements through discussion. Our analysis shows that failures partition into two distinct categories, classified based on observable trace characteristics \emph{before} any intervention.

\begin{figure}[t]
    \centering
    \includegraphics[width=0.95\columnwidth]{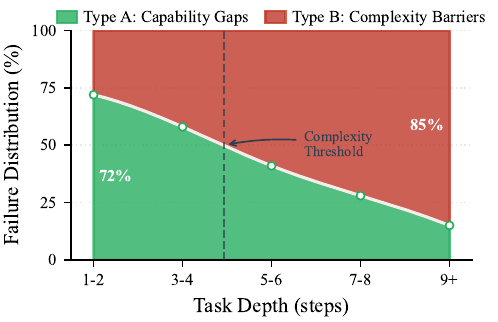}
    \caption{Failure type distribution by the task depth, measured as the number of distinct exploitation steps required for task completion. }
    \label{fig:failure-by-depth}
\end{figure}

\emph{Type A failures} (capability gaps) are identified when the trace shows the agent correctly reasons about the attack vector but fails at execution: the agent articulates the correct approach but then issues malformed commands or uses incorrect tool syntax.
For instance, an agent may correctly identify a SQL injection vulnerability (e.g., ``I will use SQL injection to extract data'') but fail because it lacks \texttt{sqlmap} or the correct documentation.
To validate this classification, we augment PentestGPT with missing tool documentation and usage instructions; XBOW completion improves from 27\% to 38\%, a 41\% relative improvement that confirms Type A failures respond to capability engineering as predicted.

\emph{Type B failures} (complexity barriers) are identified when the trace shows the agent possesses adequate tools and knowledge (evidenced by successful tool invocations earlier in the session) but fails to navigate the task space effectively.
We identify three recurring patterns from trace analysis.
\emph{Context forgetting} occurs when credentials discovered during reconnaissance are lost by the time exploitation begins, forcing redundant discovery or causing authentication failures.
\emph{Premature commitment} occurs when agents dive deep into a single attack path without adequate reconnaissance, missing easier alternatives.
\emph{Exploration-exploitation imbalance} is the inverse: exhaustive reconnaissance that never transitions to exploitation, accumulating information without acting on it.
These issues cascade into chain errors: agents complete individual attack stages successfully but fail to integrate them into coherent attack chains, losing state between phases.

\begin{table}[t]
\centering
\caption{Failure mode analysis (200 traces). Type A failures resolve with tooling; Type B persist regardless.}
\label{tab:failure-types}
\small
\begin{tabular}{lcc}
\toprule
\textbf{Failure Category} & \textbf{Freq. (\%)} & \textbf{Tooling?} \\
\midrule
\multicolumn{3}{l}{\textit{Type A: Capability Gaps (42\% total)}} \\
\quad Missing tool / Incorrect syntax & 26 & \checkmark \\
\quad Output parsing / Knowledge gap & 16 & \checkmark \\
\midrule
\multicolumn{3}{l}{\textit{Type B: Complexity Barriers (58\% total)}} \\
\quad Context forgetting & 18 & -- \\
\quad Premature commitment & 16 & -- \\
\quad Exploration-exploitation imbalance & 12 & -- \\
\quad Multi-step chain failures & 12 & -- \\
\bottomrule
\end{tabular}
\end{table}

The distribution of failure types varies systematically with task complexity.
On XBOW, where tasks typically require 1--3 steps, Type A failures dominate (68\% of failures resolve with improved tooling).
On GOAD, where successful attacks require chaining 5--10 steps across multiple hosts, Type B failures dominate (79\% of failures persist regardless of tooling improvements).
Figure~\ref{fig:failure-by-depth} visualizes this relationship: Type A failures concentrate in short-horizon tasks while Type B failures dominate in task depth beyond 5 steps. Table~\ref{tab:failure-types} summarizes the failure mode distribution.

\begin{tcolorbox}[left=1mm, right=1mm, top=0.5mm, bottom=0.5mm, arc=1mm]
\textbf{Finding 2:} Failures partition into (a) \emph{Type A: capability gaps}, i.e., missing tools and knowledge addressable through engineering, and (b) \emph{Type B: complexity barriers}, i.e., search strategy and state management failures that persist despite adequate capabilities. These two categories require different solutions.
\end{tcolorbox}

\subsection{Analysis and Design Implications}
\label{sec:study-analysis}


We now present further analysis and design implications.

\subsubsection{Root Cause: Missing Difficulty Assessment}
Type B failures share a common root cause: agents cannot distinguish tractable from intractable tasks in real time.
\emph{Premature commitment} occurs because agents cannot estimate whether a path requires 3 or 30 steps; without this estimate, they persist on unproductive branches indefinitely.
\emph{Exploration-exploitation imbalance} occurs because agents lack metrics for when reconnaissance is sufficient; they cannot determine whether gathered evidence justifies transitioning to exploitation.
\emph{Chain failures} occur partly because agents cannot assess whether their accumulated context remains adequate for the current task; critical information may have been lost or degraded without the agent's awareness.
For example, context forgetting occurs because agents lack difficulty metrics: without tracking context load, they cannot predict when accumulated history will overwhelm the model's effective memory, leading to silent degradation of reasoning quality.

What would difficulty assessment require in practice?
We identify four measurable dimensions: \emph{horizon estimation} (remaining steps to goal), \emph{evidence confidence} (certainty about current state), \emph{context load} (fraction of context window consumed), and \emph{historical success} (past performance on similar branches).
These dimensions are measurable during execution, unlike abstract ``difficulty'' which is only knowable post-hoc.
An agent that tracks these signals can decide when to persist, when to pivot, and when to prune.

\begin{figure*}[t]
    \centering
    \includegraphics[width=\textwidth]{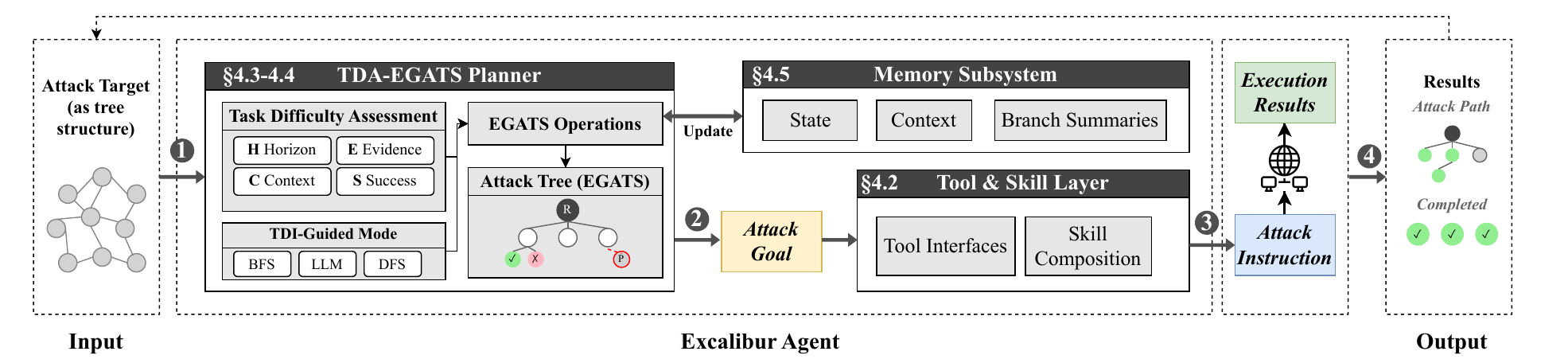}
    \caption{\tool architecture. The TDA-EGATS Planner addresses Type B failures through difficulty-aware tree search with Upper Confidence Bound (UCB) selection, TDI-guided mode switching, and evidence-based pruning. The Tool \& Skill Layer addresses Type A failures through typed tool interfaces and RAG-augmented knowledge.  The Memory Subsystem maintains structured state and enables selective context injection based on tree position.}
    \label{fig:architecture}
\end{figure*}

Current systems uniformly lack this capability.
PentestGPT's Penetration Testing Tree (PTT) tracks attack structure but provides no difficulty metrics to guide search.
AutoPT's Pentesting State Machine (PSM) enforces phase transitions but does not assess path complexity.
TermiAgent's memory tree improves context management but does not inform exploration-exploitation decisions.
None of these systems can answer the question that matters most: \emph{is this path worth pursuing?}

\subsubsection{Design Implications}
Our analysis points to a two-part strategy for advancing LLM-based penetration testing.
\emph{Eliminating Type A failures} requires comprehensive tool interfaces with typed schemas, RAG systems for exploit documentation and Common Vulnerabilities and Exposures (CVE) databases, and standardized execution environments.
This is tedious engineering work, but it produces predictable returns: each tool added directly enables new attack capabilities.

\emph{Addressing Type B failures} requires a different approach: real-time difficulty estimation, principled exploration-exploitation decisions guided by the estimates, active pruning of intractable branches to prevent search collapse, and state maintenance external to conversation context to prevent information loss.
These requirements suggest tree-based search algorithms to maintain state explicitly rather than relying on LLM's context window.

Neither approach alone is sufficient.
Capability engineering yields strong short-horizon performance but fails on complex tasks where navigation becomes the bottleneck.
Planning innovation without adequate tooling produces agents that reason well but cannot execute.
Effective systems must address both failure categories simultaneously, and in particular, agents need the ability to assess task difficulty in real time to avoid exploration-exploitation imbalance and chain failures.


\section{Design of \tool}
\label{sec:methodology}


\subsection{Overview}
We present \tool, designed around the analysis in \S\ref{sec:study-analysis} to address both failure categories through dedicated architectural components.
Figure~\ref{fig:architecture} provides its architectural overview.
\tool{} is a \textit{single-agent} system that communicates with the environment consistently, operating over different components to complete penetration testing. It consists of the following modules:
(1) A \emph{Tool and Skill Layer} that eliminates Type A failures through structured tool interfaces and knowledge augmentation (\S\ref{sec:tool-layer}).
(2) A \emph{Task Difficulty Assessment} (TDA) mechanism that estimates tractability in real time (\S\ref{sec:tda}), integrated into an \emph{Evidence-Guided Attack Tree Search} (EGATS) algorithm that replaces the traditional PTT structure for exploration-exploitation decisions (\S\ref{sec:egats}).
(3) A \emph{Memory Subsystem} that maintains state across attack phases to prevent context forgetting (\S\ref{sec:memory}).

Given a target, \tool{} \ding{182} initializes an attack tree with the target as the root node.
At each step, the EGATS planner consults the TDA module to select the current attack goal and updates the memory subsystem to preserve context.
\ding{183} The selected goal is translated into concrete actions via the Tool and Skill Layer, and \ding{184} the resulting commands are executed in the test environment.
\ding{185} Execution results are parsed and incorporated back into the attack tree and memory state, feeding into subsequent planning iterations until the penetration testing process terminates. Below we detail each component.

\subsection{Tool and Skill Layer}
\label{sec:tool-layer}

Type A failures arise not from fundamental capability limitations, but from inconsistent tool usage: LLMs invoke security tools with incorrect parameters, misparse outputs, or lack domain knowledge about tool capabilities.
Rather than proposing novel techniques, the Tool and Skill Layer represents careful engineering to ensure LLM agents interact with security tools consistently and reliably.
We build on established concepts of Agent Skills~\cite{anthropic2024skills} from Anthropic (typed interfaces, skill composition, and retrieval-augmented generation), adapting them to penetration testing where tool reliability directly determines attack success.

\noindent\textbf{Typed Tool Interfaces.}
Each security tool is exposed through a typed interface specifying input schema (parameters with types, defaults, and validation rules), output schema (structured representation parsed from command output), and pre/postconditions (required state before invocation and expected effects after completion).
The LLM receives explicit documentation rather than relying on parametric knowledge. Input validation catches errors before execution, and structured outputs eliminate parsing ambiguity.
We implement interfaces for 38 tools across six categories: reconnaissance, web exploitation, network exploitation, credential attacks, Active Directory attacks, and privilege escalation.
Appendix~\ref{sec:appendix-tools} provides the complete tool list we integrate.

\noindent\textbf{Skill Composition.}
Beyond individual tools, \emph{skills} compose multiple tool invocations into higher-level attack capabilities that encode expert knowledge about common attack patterns.
Skills provide fallback logic so that when a preferred tool fails, the system can try alternatives automatically. They also aggregate results from multiple tools into coherent findings and encode multi-step attack patterns that reflect how human testers chain operations.

\noindent\textbf{Knowledge Augmentation.}
The layer integrates a RAG system containing tool documentation, an exploit database (CVE descriptions indexed by service version), and attack playbooks (step-by-step procedures for common patterns such as Kerberoasting, AS-REP roasting, and pass-the-hash).
The knowledge base contains only generic attack techniques from public security resources (MITRE ATT\&CK, OWASP, tool documentation); it excludes CTF writeups, HTB walkthroughs, or benchmark-specific solutions to prevent data leakage in evaluation.
When the agent encounters an unfamiliar service or vulnerability class, relevant documentation is retrieved and injected into context automatically.

These three mechanisms together provide a unified, reliable interface between LLM agents and security tools.
None of these techniques is novel in isolation; their contribution lies in the combination, which minimizes tool invocation errors that otherwise cascade into attack failures.
Our ablation study (\S\ref{sec:eval-tool}) shows that this engineering effort yields substantial gains on capability-limited tasks: the Tool Layer alone improves XBOW completion by 14\% (from 54\% to 68\%), allowing agents to focus their reasoning on the harder problems of planning and strategy.

\subsection{Task Difficulty Assessment (TDA)}
\label{sec:tda}

Our analysis in \S\ref{sec:study-analysis} identifies the inability to assess task difficulty as the root cause of Type B failures.
Premature commitment occurs because agents cannot estimate whether a path requires 3 or 30 steps. Exploration-exploitation imbalance occurs because agents have no metric for when reconnaissance is sufficient. Chain failures occur because agents cannot judge whether accumulated context is adequate for the current task.

Human penetration testers face the same problem: they do not know task difficulty \emph{a priori}.
Instead, they estimate difficulty from signals that accumulate during execution, such as the number of failed attempts on a path, the quality of evidence gathered so far, and intuitions about remaining work.
An experienced tester who has tried five exploits without success knows to try a different approach; one who has confirmed a vulnerable service version knows to commit to exploitation.
TDA operationalizes this reasoning for LLM agents through four measurable dimensions, with context window consumption added as a signal unique to language models.

\subsubsection{TDA Dimensions}

TDA computes difficulty along four dimensions grounded in quantities measurable during execution.

\noindent\textbf{Horizon Estimation ($H$).} We estimate the number of remaining steps to reach the goal from the current position, normalized across active branches.
A pilot study on 50 traces from an independent GOAD deployment (using GPT-4o, separate from evaluation) shows that while absolute estimates have poor calibration (MAE of 4.2 steps), rank correlation is strong (Spearman's $\rho = 0.71$, $p < 0.001$).
The TDI formula therefore uses $\hat{H}$, the \emph{normalized} horizon estimate (min-max scaled across active branches), converting absolute estimates into relative rankings where LLM judgment is reliable.

\noindent\textbf{Historical Success Rate ($S$).}
The Laplace-smoothed success rate on the current branch captures learning from failed attempts.
Low values indicate repeated failures, suggesting that the current path is likely intractable.
This dimension directly addresses \emph{premature commitment}: agents learn to abandon unproductive paths rather than persisting indefinitely.

\noindent\textbf{Context Load ($C$).}
The fraction of context window consumed, directly measurable from token counts.
LLM performance degrades as context fills: retrieval accuracy drops, earlier information is forgotten, and reasoning quality declines~\cite{liu2024lost}.
We define an \emph{ideal working window} of 40\% of the model's context capacity, based on a controlled study showing consistent accuracy degradation beyond this point (94\% $\to$ 78\% at 60\% load, 61\% at 80\%; see \appref{sec:appendix-context}).
Beyond this threshold, context pruning becomes necessary to preserve reasoning quality.
This dimension addresses \emph{context forgetting}: by tracking context load, the system detects when accumulated history threatens to overwhelm the model's effective memory.

\noindent\textbf{Evidence Confidence ($E$).}
The mean confidence score across the path from root to current node, computed from evidence categories at each node.
We assign scores based on evidence type: verified exploits and valid credentials receive 1.0, confirmed vulnerabilities with available exploits receive 0.8, plausible hypotheses (version-matched vulnerabilities, misconfigurations) receive 0.5, and speculative hypotheses receive 0.3.
Tool outputs are parsed to determine evidence type: successful authentication or shell access indicates verified evidence, vulnerability scanner confirmations with CVE matches indicate confirmed vulnerabilities, and service version matches against exploit databases indicate plausible hypotheses.
\appref{sec:appendix-evidence} details the complete scoring rubric.
This dimension addresses \emph{exploration-exploitation imbalance}: high confidence signals readiness to exploit, while low confidence signals the need for more reconnaissance.

\subsubsection{Task Difficulty Index}

TDA combines the above 4 dimensions into a Task Difficulty Index (TDI):
\begin{equation}
\text{TDI} = w_H \cdot \hat{H} + w_E \cdot (1 - E) + w_C \cdot C + w_S \cdot (1 - S)
\label{eq:tdi}
\end{equation}
where $\hat{H}$ is the normalized horizon estimate and all weights sum to 1.
Higher TDI indicates greater difficulty.
We set $w_H = w_E = 0.3$ and $w_C = w_S = 0.2$ based on grid search over a validation set of 30 execution traces from HTB machines not included in the PentestGPT benchmark (retired machines from 2022--2023, predating our evaluation set).
We test 256 configurations with each weight in $\{0.1, 0.2, 0.3, 0.4\}$ constrained to sum to 1.0; task completion varies within $\pm$3\% across configurations where all weights remain in $[0.1, 0.4]$, indicating that the approach is not sensitive to precise weight selection.

TDI guides three operational decisions.
\emph{Mode selection:} high TDI ($> \theta_{\text{explore}} = 0.6$) triggers reconnaissance (BFS) to gather more information before committing; low TDI ($< \theta_{\text{exploit}} = 0.3$) triggers exploitation (DFS).
For intermediate values ($0.3 \leq \text{TDI} \leq 0.6$), the system invokes \textsc{LLMDecide}: the LLM receives the current node state, TDI value, and individual dimension scores ($H$, $S$, $C$, $E$), then selects between reconnaissance and exploitation with a brief justification.
This design acknowledges that intermediate difficulty may warrant either approach depending on context the TDI formula cannot fully capture. For instance, a moderately difficult path with high evidence confidence may warrant exploitation, while one with low confidence benefits from further reconnaissance.
\emph{Branch prioritization:} TDI ranks paths beyond promise scores alone, since two branches with similar promise may differ substantially in tractability based on horizon and success history.
\emph{Pruning:} branches with persistently high TDI ($> \theta_{\text{prune}} = 0.8$) after $k_{\min} = 3$ attempts are pruned to prevent the search from collapsing into unproductive regions.
These thresholds are derived through grid search on the same validation set used for TDI weights. \appref{sec:appendix-parameters} presents sensitivity analysis showing robustness across threshold ranges.

\subsection{Evidence-Guided Attack Tree Search (EGATS)}
\label{sec:egats}

EGATS integrates TDA into a tree-based search framework, adapting Monte Carlo Tree Search (MCTS)~\cite{coulom2006mcts,kocsis2006ucb} to penetration testing.
EGATS differs from standard MCTS in three ways: it explicitly separates reconnaissance (BFS) and exploitation (TDI-guided) phases, it replaces simulation-based value estimates with TDA-based difficulty assessment, and it prunes intractable branches based on evidence.

\subsubsection{Attack Tree Structure}

EGATS maintains an Attack Tree $\mathcal{T} = (V, E, \phi, \psi, \delta)$ where $V$ contains nodes representing attack states, $E$ contains edges representing actions, $\phi: V \to [0,1]$ assigns promise scores, $\psi: V \to \mathcal{S}$ maps nodes to state snapshots, and $\delta: V \to [0,1]$ assigns TDI scores.
Nodes are categorized as \emph{observation} (discovered facts), \emph{hypothesis} (untested attack possibilities), or \emph{action} (executed steps with outcomes).

The \emph{promise score} $\phi(n)$ estimates the likelihood that node $n$ leads to successful exploitation.
For hypothesis nodes, promise is initialized via LLM assessment of vulnerability severity, exploit availability, and prerequisite satisfaction; the model estimates success probability given current evidence.
For action nodes, promise is updated based on execution outcomes: successful actions propagate increased promise to ancestor nodes, while failures decrease promise along the path.
After action $a$ with outcome $o \in \{success, partial, failure\}$, we update $\phi(n) \gets \alpha \cdot \phi(n) + (1-\alpha) \cdot r(o)$ where $r(success)=1.0$, $r(partial)=0.5$, $r(failure)=0.1$, and $\alpha=0.7$ controls the learning rate.
Through this backpropagation, branches with consistent successes accumulate high promise while repeatedly failing branches see diminishing scores.

Unlike PentestGPT's text-based PTT, EGATS maintains structure externally via algorithmic operations, which prevents corruption and enables systematic search guidance.
Table~\ref{tab:search-comparison} compares EGATS with related approaches.

\begin{table}[t]
\centering
\caption{Search strategy comparison. EGATS is the only approach that combines external structure, evidence-based pruning, and TDA-guided mode selection.}
\label{tab:search-comparison}
\small
\begin{tabular}{lcccc}
\toprule
\textbf{Approach} & \textbf{Structure} & \textbf{Pruning} & \textbf{Difficulty} & \textbf{TDA} \\
\midrule
ReAct & None & -- & -- & -- \\
PTT~\cite{deng2024pentestgpt} & Tree (text) & Manual & -- & -- \\
PSM~\cite{autopt2024} & Finite state machine & -- & -- & -- \\
PMT~\cite{termibench2024} & Tree & -- & -- & -- \\
\textbf{EGATS} & Tree (ext.) & Evidence & \checkmark & \checkmark \\
\bottomrule
\end{tabular}
\end{table}

\subsubsection{The EGATS Algorithm}

Algorithm~\ref{alg:egats} presents the TDA-guided search procedure.
\textsc{SelectNode} uses UCB to balance exploitation and exploration:
\begin{equation}
\text{UCB}(n) = \phi(n) + c \sqrt{\frac{\ln N}{N_n}} - \lambda \cdot \delta(n)
\label{eq:ucb}
\end{equation}
where $\phi(n)$ is the promise score, $N$ is total actions, $N_n$ is actions on node $n$'s subtree, $c = \sqrt{2}$ is the exploration constant, and the $-\lambda \cdot \delta(n)$ term penalizes high-difficulty nodes ($\lambda = 0.5$, validated via grid search; see \appref{sec:appendix-parameters}).

After selection, EGATS computes TDI and switches between BFS (reconnaissance) and DFS (exploitation) based on the thresholds described above.
Evidence backpropagates after each action, updating promise scores and TDI along affected paths.
When exploitation succeeds, \emph{pivot spawning} is triggered: the compromised host becomes a new subtree root, and discovered credentials propagate to relevant hypothesis nodes elsewhere in the tree.

Pruning removes branches when TDI exceeds 0.8 after three attempts, which prevents infinite loops on intractable paths.
To avoid premature pruning, a credential propagation mechanism re-evaluates pruned branches when new credentials are discovered that may satisfy their preconditions.

\begin{algorithm}[t]
\caption{TDA-Guided Attack Tree Search}
\label{alg:egats}
\begin{algorithmic}[1]
\Require Target $T$, budget $B$
\Ensure Attack tree $\mathcal{T}$, compromised hosts $C$
\State $\mathcal{T} \gets$ \Call{InitTree}{$T$}
\While{$B > 0$ \textbf{and not} \Call{GoalReached}{}}
    \State $n \gets$ \Call{SelectNode}{$\mathcal{T}$} \Comment{UCB selection}
    \State $\text{TDI}_n \gets$ \Call{ComputeTDI}{$n$}
    \If{$\text{TDI}_n > \theta_{\text{explore}}$}
        \State \Call{ExecuteRecon}{$n$}; \Call{ExpandTree}{$\mathcal{T}$, $n$}
    \ElsIf{$\text{TDI}_n < \theta_{\text{exploit}}$}
        \State $result \gets$ \Call{ExecuteExploit}{$n$}
        \State \Call{BackpropagateEvidence}{$\mathcal{T}$, $n$, $result$}
        \If{$result.success$} \Call{SpawnPivot}{$\mathcal{T}$, $result.host$}
        \EndIf
    \Else
        \State \Call{LLMDecide}{$n$, $\text{TDI}_n$}
    \EndIf
    \If{$\delta(n) > \theta_{\text{prune}}$ \textbf{and} $N_n > k_{\min}$}
        \State \Call{PruneBranch}{$\mathcal{T}$, $n$}
    \EndIf
    \State $B \gets B - 1$
\EndWhile
\end{algorithmic}
\end{algorithm}

\subsection{Memory Subsystem}
\label{sec:memory}

Long-context forgetting is a primary cause of Type B failures (\S\ref{sec:study-findings}).
The Memory Subsystem addresses this with a hybrid architecture that separates persistent state from conversational context, and integrates with TDA via the context load dimension.

A \emph{State Store} maintains a structured database of discovered facts independent of conversation context.
The store tracks five entity types: hosts (IP addresses, hostnames, OS fingerprints), services (ports, versions, configurations), credentials (usernames, passwords, hashes, tickets), sessions (active shells, tunnels, pivots), and vulnerabilities (CVE identifiers, exploitation status, prerequisites).
Each entry is timestamped and linked to its discovery node in the attack tree, which enables provenance tracking and ensures facts persist regardless of conversation length.
The State Store also supports accurate TDA context load computation by providing ground truth about what information the agent ``knows'' versus what must be re-derived from context.

\emph{Selective context injection} replaces full history maintenance.
When operating on node $n$, context is assembled from: path context (the sequence of actions from root to $n$), a node state snapshot (complete state at $n$ including all relevant entity relationships), target-relevant facts (entries from State Store pertaining to $n$'s target host or service), and sibling branch summaries (compressed representations of parallel exploration paths).
As context load approaches the ideal working window threshold (40\%), less-relevant context is progressively compressed using LLM-generated summaries. Beyond 70\%, aggressive pruning removes older path segments while preserving findings to prevent performance degradation.

\emph{Branch summaries} compress detailed execution history when switching branches.
Each summary preserves the current status (active, pruned, completed), findings (discovered credentials, confirmed vulnerabilities), TDI at time of suspension, and recommended next actions.
TDI is stored with each summary to inform revisit decisions: when new credentials are discovered elsewhere in the tree, branches with matching preconditions and previously high TDI are re-evaluated for potential reactivation.

\section{Evaluation}\label{sec:evaluation}

We assess the performance of \tool{} through four research questions:
\begin{itemize}
[leftmargin=*,noitemsep,topsep=5pt,parsep=0pt,partopsep=0pt]
    \item \textbf{RQ1}: Does \tool{} outperform existing systems across different penetration testing scenarios?
    \item \textbf{RQ2}: What is the contribution of the each designed architectural component?
    \item \textbf{RQ3}: How does TDA-EGATS change the agent's attack strategy compared to prior approaches?
    \item \textbf{RQ4}: Can \tool{} be practically deployed for real-world penetration testing?
\end{itemize}

\begin{table*}[t]
\centering
\caption{Performance comparison across systems, models, and benchmarks. Each model column is split into non-thinking (--) and thinking (T) modes. XBOW: task completion (\%); PentestGPT Benchmark: machines rooted (/13); GOAD: hosts compromised (/5). Best results per column in \textbf{bold}. All results report mean across 3 trials; variance $\pm$2--3\% on XBOW, $\pm$1 machine on PentestGPT-Ben.}
\label{tab:main-results}
\small
\setlength{\tabcolsep}{3pt}
\begin{tabular}{l|cc|cc|cc|cc|cc|cc|cc|cc|cc}
\toprule
& \multicolumn{6}{c|}{\textbf{XBOW (104 tasks)}} & \multicolumn{6}{c|}{\textbf{PentestGPT-Ben (13 machines)}} & \multicolumn{6}{c}{\textbf{GOAD (5 hosts)}} \\
& \multicolumn{2}{c|}{GPT-5.2} & \multicolumn{2}{c|}{Opus 4.5} & \multicolumn{2}{c|}{Gemini 3} & \multicolumn{2}{c|}{GPT-5.2} & \multicolumn{2}{c|}{Opus 4.5} & \multicolumn{2}{c|}{Gemini 3} & \multicolumn{2}{c|}{GPT-5.2} & \multicolumn{2}{c|}{Opus 4.5} & \multicolumn{2}{c}{Gemini 3} \\
\textbf{System} & -- & T & -- & T & -- & T & -- & T & -- & T & -- & T & -- & T & -- & T & -- & T \\
\midrule
PentestGPT & 45 & 53 & 47 & 54 & 41 & 48 & 7 & 8 & 6 & 7 & 6 & 7 & 1 & 1 & 1 & 2 & 1 & 1 \\
AutoPT & 43 & 50 & 44 & 51 & 38 & 45 & 6 & 7 & 7 & 8 & 5 & 6 & 1 & 1 & 0 & 1 & 1 & 1 \\
PentestAgent & 52 & 61 & 54 & 60 & 46 & 54 & 8 & 9 & 7 & 9 & 7 & 8 & 1 & 2 & 2 & 2 & 1 & 1 \\
VulnBot & 48 & 56 & 50 & 58 & 43 & 51 & 8 & 9 & 8 & 9 & 6 & 8 & 2 & 2 & 1 & 2 & 1 & 2 \\
\midrule
\textbf{\tool{}} & \textbf{76} & \textbf{85} & \textbf{81} & \textbf{91} & \textbf{76} & \textbf{79} & \textbf{11} & \textbf{12} & \textbf{10} & \textbf{12} & \textbf{10} & \textbf{11} & \textbf{3} & \textbf{4} & \textbf{3} & \textbf{4} & \textbf{3} & \textbf{3} \\
\bottomrule
\end{tabular}
\end{table*}

\subsection{Experimental Setup}

\tool is implemented in Python ($\sim$8,500 lines), with the Tool Layer, TDA-EGATS Planner, and Memory Subsystem as separate modules. The implementation is open-sourced~\cite{excalibur2025artifact}.
Following the evaluation methodology in Section~\ref{sec:study-survey}, we evaluate \tool{} on three benchmarks of increasing complexity.
\textbf{XBOW}~\cite{xbow2024} comprises 104 CTF-style web security challenges covering SQL injection, XSS, authentication bypass, and file inclusion; these short-horizon tasks isolate Type A failures where tool usage determines success.
The \textbf{PentestGPT Benchmark}~\cite{deng2024pentestgpt} consists of 13 machines from HTB and VulnHub, requiring end-to-end penetration testing from reconnaissance through privilege escalation to root access.
Difficulty ranges from Easy to Hard, with 9--22 subtasks per machine, representing realistic scenarios that demand multi-step attack chains.
\textbf{GOAD}~\cite{goad2024} provides a 5-host multi-domain Active Directory environment requiring credential harvesting, Kerberoasting, lateral movement, and domain escalation, complex enterprise scenarios dominated by Type B failures.

We compare against four baseline systems: PentestGPT v1.0~\cite{deng2024pentestgpt}, AutoPT~\cite{autopt2024}, PentestAgent~\cite{chen2024pentestagent}, and VulnBot~\cite{vulnbot2025}.
We exclude Cochise~\cite{happe2025cochise} from this comparison because its AD-specialized architecture creates an uneven evaluation as shown in Section~\ref{sec:study-findings}. 
Baseline systems use their original tool invocation mechanisms to reflect realistic deployment comparisons; reported improvements therefore reflect both tool integration and architectural contributions.
To isolate architectural contributions from model capabilities, all systems are evaluated with three frontier models: GPT-5.2, Claude-Opus-4.5, and Gemini-3.0-Pro.
We select these models for two reasons: (1) they represent state-of-the-art capabilities at the time of evaluation, and (2) all three support toggling between standard and thinking modes, enabling controlled comparison of extended reasoning effects.
We report task completion rate, subtask progress, and exploration metrics including branch diversity, backtrack frequency, and time-to-pivot.
We report mean performance across trials with standard deviation where variance is meaningful; for discrete outcomes (machines rooted, hosts compromised), we report best-of-three following prior work~\cite{deng2024pentestgpt,autopenbench2024} since standard deviation on small integers provides limited insight.
For XBOW's continuous completion rates, we report both headline best-of-three results and trial statistics ($\mu$: mean, $\sigma$: standard deviation across the three trials) to characterize variance.
In total, we conduct 5 systems $\times$ 118 evaluation units $\times$ 6 model configurations $\times$ 3 trials, yielding 10,620 evaluation runs at an estimated cost of \$2,760 USD in API tokens (Table~\ref{tab:cost} reports \tool-specific costs).

\subsection{RQ1: Overall Performance}

Table~\ref{tab:main-results} shows the performance comparison across all system-model-benchmark combinations, with consistent patterns that align with our Type A/B failure framework.

On XBOW, \tool{} achieves 91\% task completion (best-of-3; $\mu$=89\%, $\sigma$=2.1\%) with Opus 4.5 thinking mode, a 49\% relative improvement over the best baseline (PentestAgent at 61\%, $\mu$=59\%, $\sigma$=1.8\%).
With GPT-5.2 thinking, \tool{} achieves 85\% ($\mu$=83\%, $\sigma$=2.4\%) compared to 61\% for PentestAgent.
Even comparing means, the gap (89\% vs.\ 59\%) exceeds 15 standard deviations, confirming robust architectural differences: the Tool Layer eliminates Type A failures while TDA-EGATS prevents trial-and-error loops that consume baseline attempts.
Thinking mode provides 6--10 point improvements across all systems and configurations but does not close the architectural gap.

The PentestGPT benchmark shows larger architectural differences.
\tool{} roots 12 of 13 machines with both GPT-5.2 and Opus 4.5 thinking (consistent across all three trials), compared to 9 for the best baseline (VulnBot), a 33\% relative improvement.
\tool{} solves both Hard-rated machines (Joker and Falafel) where baseline systems became ``stuck at initial steps,'' and also completes machines that require non-obvious attack chains.
The improvement concentrates in machines requiring non-linear attack paths: while baseline PTT structures lead to premature commitment on initial hypotheses, TDA-EGATS enables strategic backtracking when evidence confidence drops, allowing the agent to discover alternative attack vectors.
Thinking mode amplifies architectural differences: \tool{} gains 1--2 machines from thinking, achieving near-complete coverage, while baselines gain only 1 machine each but plateau at 9.

GOAD shows the largest improvement.
\tool{} compromises 4 of 5 hosts with GPT-5.2 and Opus 4.5 thinking (4 hosts in all three trials; the same four hosts each time) versus at most 2 for baselines---doubling the compromise rate (80\% vs.\ 40\%).
This pattern holds consistently across all three models and both reasoning modes (even Gemini 3 achieves 3 hosts vs.\ 1--2 for baselines), indicating a robust architectural effect.
Baselines achieve initial foothold but fail to progress through lateral movement; \tool{} executes coherent multi-host attack chains using the Memory Subsystem for credential persistence and TDA for exploration guidance.

\subsection{RQ2: Ablation Study}
\label{sec:eval-tool}

\begin{figure}[t]
    \centering
    \includegraphics[width=\columnwidth]{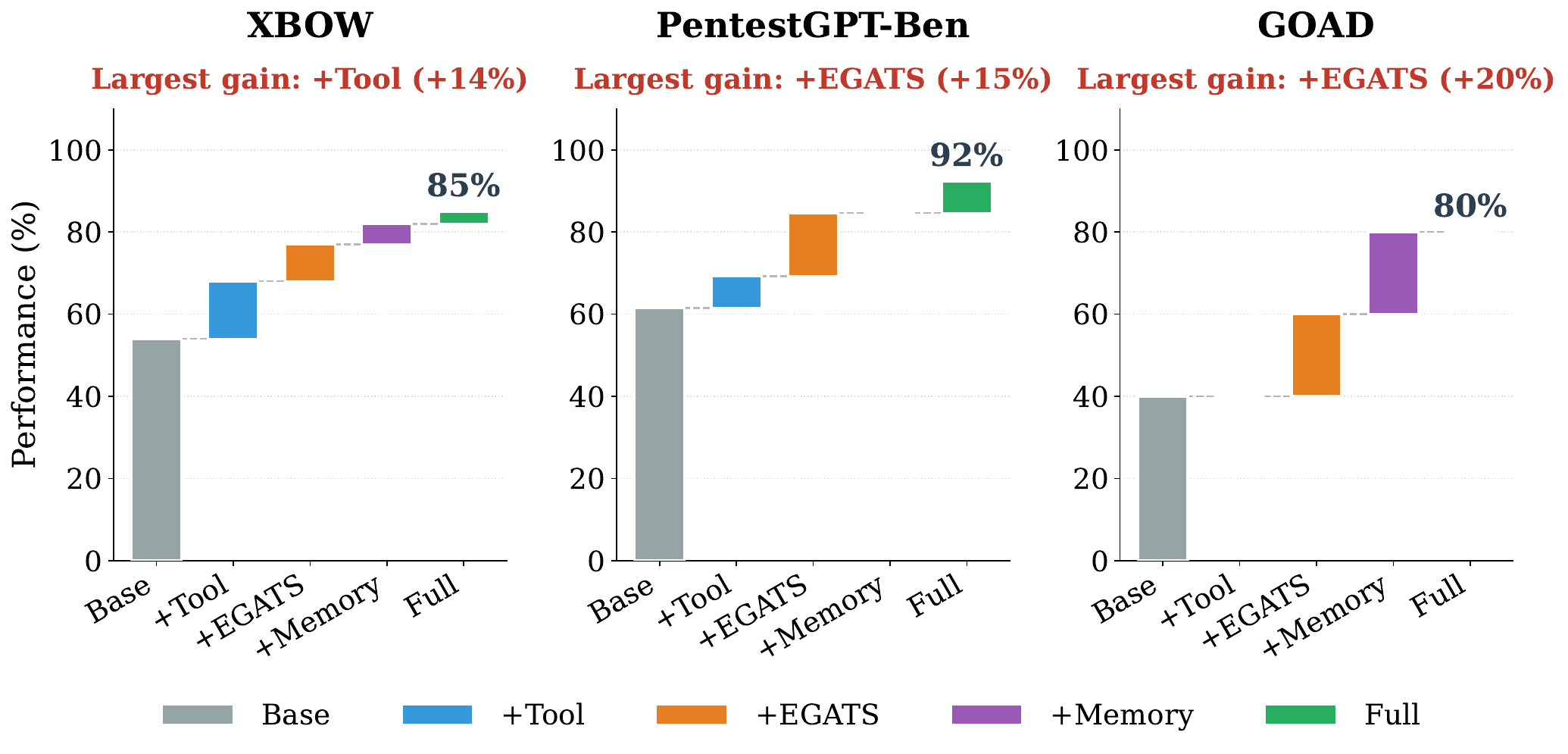}
    \caption{Ablation study across benchmarks (GPT-5.2 thinking). Performance is normalized to percentage scale. }
    \label{fig:ablation}
\end{figure}

To isolate each component's contribution, we evaluate system variants with individual components disabled.
Table~\ref{tab:ablation} presents results using GPT-5.2 thinking mode; the base configuration uses raw shell access with reactive prompting and sliding-window context management.
Figure~\ref{fig:ablation} visualizes component contributions across all model configurations.

\begin{table}[t]
\centering
\caption{Ablation study results (GPT-5.2 thinking). Base: raw shell access with reactive prompting. Each row adds a component cumulatively.}
\label{tab:ablation}
\small
\begin{tabular}{lccc}
\toprule
\textbf{Configuration} & \textbf{XBOW} & \textbf{Pentest-Ben} & \textbf{GOAD} \\
\midrule
Base & 54 & 8 & 2 \\
+ Tool Layer & 68 & 9 & 2 \\
+ TDA-EGATS & 77 & 11 & 3 \\
+ Memory (Full) & \textbf{85} & \textbf{12} & \textbf{4} \\
\bottomrule
\end{tabular}
\end{table}

The results align with our Type A/B failure framework.
The Tool Layer provides the largest improvement on XBOW (+14 points, from 54 to 68), consistent with CTF failures being predominantly engineering problems addressable through better tooling.
The Tool Layer alone yields zero improvement on GOAD (remaining at 2 hosts), where planning rather than capability determines success.

TDA-EGATS adds further gains: +9 points on XBOW (from 68 to 77) through reduced trial-and-error, +2 machines on the PentestGPT benchmark (from 9 to 11), and +1 host on GOAD (from 2 to 3).
These gains span both Type A failures (via more efficient search) and Type B failures (via principled exploration-exploitation).
The Memory Subsystem contributes across all benchmarks: +8 points on XBOW (from 77 to 85), +1 machine on the PentestGPT benchmark (from 11 to 12), and +1 host on GOAD (from 3 to 4).
The GOAD improvement is worth noting separately: extended attack campaigns cause context forgetting in systems without explicit state management, and Memory enables the credential persistence required for the fourth compromise.

\subsection{RQ3: Strategy Analysis}
\label{sec:eval-tda}

Beyond aggregate performance, we analyze how TDA-EGATS changes the agent's attack strategy compared to PentestGPT's PTT-based approach.

\subsubsection{Search Behavior}

Table~\ref{tab:strategy} compares exploration patterns across the PentestGPT benchmark.
The metrics show qualitatively different search behaviors between the two systems.

\begin{table}[t]
\centering
\caption{Search behavior comparison on the PentestGPT benchmark (mean across 13 machines).}
\label{tab:strategy}
\small
\begin{tabular}{lcc}
\toprule
\textbf{Metric} & \textbf{PentestGPT} & \textbf{\tool{}} \\
\midrule
Branches explored & 3.2 & 7.8 \\
Backtrack rate (\%) & 8 & 34 \\
Avg. depth before pivot & 12.4 & 5.1 \\
Successful pivots & 0.4 & 2.6 \\
Pruned branches & -- & 4.2 \\
\bottomrule
\end{tabular}
\end{table}

PentestGPT follows a deep-first pattern: it explores fewer branches (3.2 vs.\ 7.8) but commits to each for longer (average depth 12.4 steps before pivoting vs.\ 5.1 for \tool{}), reflecting the premature commitment failure mode where agents persist on initial hypotheses without signals to recognize intractability.

\tool{} with TDA-EGATS follows an adaptive pattern: TDI monitoring triggers backtracking when success rate drops, and evidence confidence guides exploitation timing.
The 4.2 pruned branches per machine are paths abandoned due to persistently high TDI, preventing the infinite loops observed in baseline systems.

\subsubsection{Case Study: HTB Falafel}

\begin{figure}[t]
    \centering
    \includegraphics[width=\columnwidth]{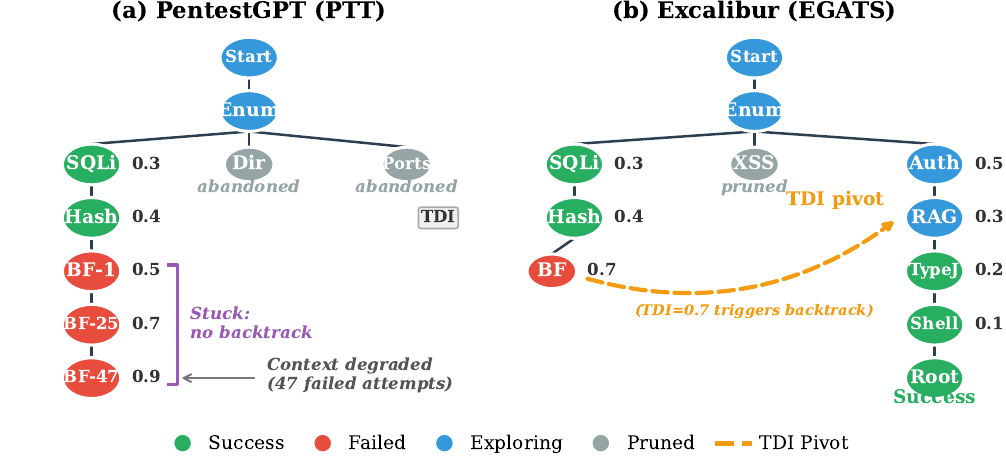}
    \caption{HTB Falafel exploration comparison. (a) PentestGPT commits to password brute-force after extracting hashes and stalls after 47 attempts. (b) \tool{}'s TDI-guided exploration discovers the type juggling bypass when hash cracking fails, then navigates the privilege escalation chain.}
    \label{fig:falafel-combined}
\end{figure}

Falafel is a Hard-rated HTB machine requiring a multi-stage attack chain that combines web exploitation, cryptographic quirks, and privilege escalation through Linux group memberships.
Figure~\ref{fig:falafel-combined} contrasts how PentestGPT and \tool{} navigate this challenge.

The attack begins with web enumeration revealing a login form that produces different error messages for valid versus invalid usernames, enabling user discovery through fuzzing.
Boolean-based blind SQL injection in the username field allows extracting password hashes from the database.
The key step is recognizing that the admin hash begins with ``0e462...'', a format that PHP's loose comparison operator (\texttt{==}) interprets as scientific notation.
Submitting the string ``240610708'' produces an MD5 hash also starting with ``0e'', causing both values to compare as zero and bypassing authentication without password cracking.
Post-authentication, a filename truncation vulnerability enables code execution: the system truncates filenames exceeding 237 characters, so uploading a file named \texttt{[232 A's].php.png} results in an executable \texttt{.php} file after truncation removes the \texttt{.png} extension.
Privilege escalation chains through three stages: database credentials in the PHP configuration yield user \texttt{moshe}; membership in the \texttt{video} group enables framebuffer capture that reveals \texttt{yossi}'s password displayed on screen; membership in the \texttt{disk} group allows reading root's files directly via \texttt{debugfs}.

PentestGPT successfully extracts the password hashes but commits to direct cracking via hashcat.
After 47 failed attempts with various wordlists and rules, context degradation prevents the model from revisiting the hash format---the type juggling vector is never considered.

\tool{}'s EGATS tree develops differently.
When hash cracking yields repeated failures, rising TDI triggers exploration of authentication alternatives.
The Knowledge Augmentation component surfaces PHP type juggling documentation when queried about hashes starting with ``0e'', enabling the bypass.
The Memory Subsystem preserves credentials discovered at each privilege escalation stage, enabling the complete chain from \texttt{www-data} through \texttt{moshe} and \texttt{yossi} to \texttt{root}.

\subsubsection{Failure Case: PlayerTwo}

To illustrate where TDA-EGATS falls short, we examine PlayerTwo, the only PentestGPT Benchmark machine \tool{} fails to compromise.
PlayerTwo requires exploiting a custom Protobuf-based game protocol with no public documentation.
\tool{} correctly identifies the service through reconnaissance and spawns hypothesis branches for protocol fuzzing.
However, TDI rises rapidly due to repeated failures (low $S$) and high horizon estimates (the LLM cannot predict steps for an unknown protocol).
After three unsuccessful fuzzing attempts, the branch is pruned correctly by TDA's design logic, since success rate indicates intractability.

This failure exposes a TDA limitation: it cannot distinguish ``difficult but tractable'' from ``novel requiring creative reasoning,'' as both present as high TDI.
When RAG retrieval finds no relevant documentation and the LLM lacks parametric knowledge, TDA's evidence-based signals provide no useful guidance.
TDA-EGATS therefore improves navigation through \emph{known} attack spaces but does not address \emph{novel} exploitation requiring genuine invention.

\subsection{RQ4: Real-World Deployment}
\label{sec:eval-realworld}

To assess practical viability, we evaluate \tool{}'s resource consumption. We further deploy it in a live competition environment to examine its real-world performance.

\subsubsection{Cost Analysis}

\begin{table}[t]
\centering
\caption{Resource consumption per task (median values, GPT-5.2 thinking).}
\label{tab:cost}
\small
\begin{tabular}{lccc}
\toprule
\textbf{Benchmark} & \textbf{LLM Calls} & \textbf{Time (min)} & \textbf{Cost (\$)} \\
\midrule
XBOW  & 12 & 3.2 & 0.18 \\
PentestGPT-Ben & 87 & 42 & 4.20 \\
GOAD & 234 & 186 & 28.50 \\
\bottomrule
\end{tabular}
\end{table}

Table~\ref{tab:cost} presents the resource consumption across benchmarks.
\tool{} requires 23\% fewer LLM calls than the baseline average on XBOW (12 vs.\ 15.6 median calls per task) due to reduced trial-and-error from structured tool interfaces, while achieving 39\% higher success rates (85\% vs.\ 61\%).
On GOAD, total calls increase by 18\% due to more thorough exploration enabled by EGATS, but this yields 2$\times$ more compromised hosts (4 vs.\ 2).
On a per-success basis, \tool{} is 1.8$\times$ more cost-effective on XBOW and 1.7$\times$ more cost-effective on GOAD: the overhead of EGATS is more than offset by the higher success rates.
A complete GOAD engagement costs approximately \$28.50 and achieves 80\% environment compromise (4 of 5 hosts), making automated penetration testing economically viable for enterprise security assessments.

\subsubsection{Live Competition Deployment}

We deployed \tool{} during HTB Season 8 (May--August 2025), a live competition with 13 newly released machines whose solutions remain unavailable until the season concludes.
This provides a direct test of real-world viability: unlike retired benchmark machines, Season machines incorporate recent CVEs and novel attack chains with no public walkthroughs.

\tool{} with Opus 4.1 completed 10 of 13 machines (76.9\%), achieving a global ranking in the top 100 out of 8,036 active participants.

\begin{table}[t]
\centering
\caption{HTB Season 8 performance by difficulty (May--August 2025). Total: 10/13 machines (76.9\%).}
\label{tab:season8}
\small
\begin{tabular}{lccc}
\toprule
\textbf{Difficulty} & \textbf{Completed} & \textbf{Total} & \textbf{Rate} \\
\midrule
Easy & 4 & 4 & 100\% \\
Medium & 4 & 4 & 100\% \\
Hard & 2 & 3 & 67\% \\
Insane & 0 & 2 & 0\% \\
\midrule
\textbf{Total} & \textbf{10} & \textbf{13} & \textbf{76.9\%} \\
\bottomrule
\end{tabular}
\end{table}

Table~\ref{tab:season8} summarizes performance by difficulty.
All four Easy machines and all four Medium machines were compromised successfully.
Among Hard machines, \tool{} completed Certificate and RustyKey but failed on Mirage.
Both Insane machines, \textit{Sorcery} and \textit{Cobblestone}, remained unsolved.
The three failures, Mirage (Hard), Sorcery (Insane), and Cobblestone (Insane), represent machines where \tool{} exhausted its search space without finding viable attack paths.
These results align with the PlayerTwo analysis (\S\ref{sec:eval-tda}): when RAG retrieval yields no relevant documentation and the underlying model lacks parametric knowledge of the target vulnerability class, TDA-EGATS cannot guide exploration effectively.

The Season 8 deployment shows that \tool{} can operate in realistic penetration testing scenarios where solutions are unknown and time-constrained.
The 100\% success rate on Easy and Medium machines suggests readiness for deployment on typical enterprise targets, while Hard and Insane failures mark the current boundaries where human expertise is still required.

\section{Discussion}
\label{sec:discussion}

\subsection{Limitations and Threats to Validity}

We discuss factors that bound the generalizability of our findings.

\noindent\textbf{Benchmark Scope.}
Our evaluation covers web security, network penetration testing, and Active Directory attacks, but omits binary exploitation, mobile security, and cloud-specific attack scenarios where different challenges may dominate.
Binary exploitation requiring precise memory layout reasoning poses distinct challenges not captured by our benchmarks.
The PentestGPT Benchmark uses retired machines with public walkthroughs, which may inflate absolute numbers through data contamination; however, TDA, EGATS, and Memory target planning challenges orthogonal to specific vulnerability knowledge and thus transfer to novel scenarios.
Real-world engagements also involve active defenses and novel vulnerability classes absent from historical benchmarks.

\noindent\textbf{Model-Specific Effects.}
We obtain results with three frontier models (GPT-5.2, Claude-Opus-4.5, Gemini-3.0-Pro).
Different model architectures show different strengths: Opus 4.5 achieves the highest XBOW performance (91\%), which suggests that our architectural contributions may interact differently across model families.
Future model generations may shift the easy/hard boundary and potentially resolve challenges we currently classify as hard.


\noindent\textbf{Baseline Fairness.}
We use published baseline code with default parameters; original authors might achieve better results through tuning, though this reflects realistic deployment scenarios.
Because baselines use their original tool invocation mechanisms, reported improvements reflect both tool integration and architectural contributions.

\noindent\textbf{Failure Analysis.}
We analyze \tool's remaining failures to characterize current boundaries.
On XBOW, the 9 failed tasks (9\%) fall into two categories: blind injection that requires extensive timing-based exfiltration (4 tasks), and multi-stage attacks that require creative payload chaining not present in our RAG corpus (5 tasks).
The single unsolved PentestGPT Benchmark machine (PlayerTwo, Hard) requires exploiting a custom protocol with no public documentation, a novel exploitation scenario that demands reasoning beyond pattern matching.
On GOAD, the fifth host (the forest root domain controller) requires a specific attack chain (PrintNightmare $\rightarrow$ DCSync) that \tool identifies but fails to execute due to token constraints.
These failures indicate that while \tool addresses Type B failures effectively, novel exploitation that requires creative reasoning remains an open problem.

\subsection{What Remains Hard}

Despite \tool's gains, three categories of irreducible Type B failures persist that better tooling, larger corpora, or improved prompting cannot resolve.

\noindent\textbf{The Creativity Barrier.}
LLMs are effective at pattern matching but struggle with out-of-distribution generalization~\cite{mirzadeh2024gsm}.
The PlayerTwo failure illustrates this gap: \tool systematically explores attack vectors yet fails because no documented exploitation pattern exists for the custom Protobuf-based protocol.
The distinction between ``difficult'' and ``novel'' matters here. Difficult tasks respond to improved search; novel tasks require reasoning capabilities that current architectures do not provide.

\noindent\textbf{The Adversarial Environment Barrier.}
Penetration testing occurs against active defenders who can exploit agent reasoning patterns~\cite{zhan2025adaptive}.
Honeypots, canary tokens, and deceptive services can poison the agent's state representation, causing it to pursue false attack paths or trigger detection.
\tool's evidence grounding protects against self-generated hallucinations but offers limited defense against environmentally-induced false beliefs: when a honeypot presents a convincing vulnerable service, the agent cannot tell whether the vulnerability is genuine or a deliberate trap.
This asymmetry favors defenders, who can study and exploit agent behavior, while agents lack the meta-awareness to recognize manipulation.

\noindent\textbf{The Temporal Scale Barrier.}
Human pentesters maintain mental models across engagements that span weeks, correlating information from separate sessions and exercising strategic patience.
EGATS improves multi-step reasoning within sessions and the Memory Subsystem preserves state, but neither addresses cross-session continuity.
Long-horizon planning is a different problem from long-context processing: it requires hierarchical abstraction, goal decomposition, and progress monitoring, none of which current transformer architectures natively support~\cite{packer2023memgpt}.

\section{Conclusion}
\label{sec:conclusion}

This paper presents a systematic analysis of LLM-based penetration testing that identifies a distinction between Type A failures (capability gaps addressable through engineering) and Type B failures (complexity barriers requiring architectural innovation).
We introduce \tool, which addresses Type A failures through a Tool and Skill Layer with typed interfaces and RAG, and addresses Type B failures via Task Difficulty Assessment (TDA) integrated into Evidence-Guided Attack Tree Search (EGATS).
\tool achieves 91\% task completion on CTF benchmarks (49\% improvement over baselines) and compromises 4 of 5 hosts on the GOAD Active Directory environment versus 2 for prior systems.
Our ablation studies show that TDA-guided exploration provides benefits beyond tree structure alone: difficulty-aware planning produces value that model improvements cannot replicate.

\clearpage
\bibliographystyle{plainurl}
\bibliography{bib}

\appendix
\section{Surveyed LLM-Based Penetration Testing Systems}
\label{sec:appendix-systems}

Table~\ref{tab:appendix-systems} presents the complete list of 28 candidate systems identified in our survey.
Systems meeting our inclusion criteria (LLM as core component, targeting penetration testing or CTF challenges, with published technical details) are marked with \checkmark.

\begin{table}[h]
\centering
\caption{Complete list of surveyed LLM-based penetration testing systems. Systems marked with \checkmark meet our inclusion criteria and are analyzed in Section~\ref{sec:study}.}
\label{tab:appendix-systems}
\small
\begin{tabular}{llcc}
\toprule
\textbf{System} & \textbf{Source} & \textbf{Year} & \textbf{Included} \\
\midrule
PentestGPT~\cite{deng2024pentestgpt} & USENIX Security & 2024 & \checkmark \\
AutoPT~\cite{autopt2024} & arXiv & 2024 & \checkmark \\
RapidPen~\cite{nakatani2025rapidpen} & arXiv & 2025 & \checkmark \\
PentestAgent~\cite{chen2024pentestagent} & arXiv & 2024 & \checkmark \\
VulnBot~\cite{vulnbot2025} & arXiv & 2025 & \checkmark \\
xOffense~\cite{xoffense2024} & arXiv & 2025 & \checkmark \\
TermiAgent~\cite{termibench2024} & arXiv & 2025 & \checkmark \\
HackSynth~\cite{muzsai2024hacksynth} & arXiv & 2024 & \checkmark \\
MAPTA~\cite{david2025mapta} & arXiv & 2025 & \checkmark \\
Cochise~\cite{happe2025cochise} & arXiv & 2025 & \checkmark \\
\midrule
\multicolumn{4}{l}{\textit{Excluded: Vulnerability detection only}} \\
VulnScanner-AI & GitHub & 2024 & \\
LLM-SecAudit & arXiv & 2024 & \\
CodeVuln & arXiv & 2024 & \\
BugHunter & RAID & 2024 & \\
AutoFuzz-LLM & CCS & 2024 & \\
\midrule
\multicolumn{4}{l}{\textit{Excluded: Commercial/no details}} \\
Pentera & Commercial & 2024 & \\
Cobalt Strike AI & Commercial & 2024 & \\
CrowdStrike Charlotte & Commercial & 2024 & \\
\midrule
\multicolumn{4}{l}{\textit{Excluded: Non-exploitation focus}} \\
CTF-Helper & arXiv & 2023 & \\
CryptoSolver & arXiv & 2024 & \\
RevEngGPT & arXiv & 2024 & \\
MalwareGPT & arXiv & 2024 & \\
ThreatGPT & arXiv & 2024 & \\
SecurityBot & GitHub & 2024 & \\
DFIR-Assistant & arXiv & 2024 & \\
IRBot & arXiv & 2025 & \\
SOC-Copilot & arXiv & 2024 & \\
VulnReport-LLM & arXiv & 2024 & \\
\bottomrule
\end{tabular}
\end{table}

\section{Tool and Skill Layer: Supported Tools}
\label{sec:appendix-tools}

Table~\ref{tab:appendix-tools} lists the 38 security tools integrated into \tool's Tool and Skill Layer.
Each tool is exposed through a typed interface specifying input parameters, output schema, and pre/postconditions.
Tool selection reflects standard penetration testing methodology and aligns with tools commonly used in professional certifications (e.g., OSCP) and real-world assessments.

\begin{table*}[t]
\centering
\caption{Security tools integrated into \tool. Each tool has a typed interface specifying input schema, output parsing, and execution constraints.}
\label{tab:appendix-tools}
\small
\begin{tabular}{llp{8.5cm}}
\toprule
\textbf{Category} & \textbf{Tool} & \textbf{Description} \\
\midrule
\multirow{8}{*}{Reconnaissance}
    & \texttt{nmap} & Network discovery, port/service scanning, OS fingerprinting \\
    & \texttt{masscan} & High-speed port scanner for large networks \\
    & \texttt{gobuster} & Directory/DNS bruteforcing for web discovery \\
    & \texttt{ffuf} & Web fuzzer for directories, parameters, vhosts \\
    & \texttt{feroxbuster} & Recursive web content discovery \\
    & \texttt{nikto} & Web server vulnerability scanner \\
    & \texttt{whatweb} & Web technology fingerprinting \\
    & \texttt{enum4linux} & SMB/Samba enumeration (users, shares, OS) \\
\midrule
\multirow{6}{*}{Web Exploitation}
    & \texttt{sqlmap} & SQL injection detection and exploitation \\
    & \texttt{burpsuite} & Web proxy for traffic interception and testing \\
    & \texttt{zap} & OWASP web vulnerability scanner \\
    & \texttt{wfuzz} & Web fuzzer for parameters and authentication \\
    & \texttt{commix} & Command injection exploitation \\
    & \texttt{nuclei} & Template-based CVE and misconfiguration scanner \\
\midrule
\multirow{7}{*}{Network Exploitation}
    & \texttt{metasploit} & Exploitation framework with pre/post-exploitation modules \\
    & \texttt{netcat} & TCP/UDP networking utility \\
    & \texttt{crackmapexec} & Windows/AD post-exploitation toolkit \\
    & \texttt{responder} & LLMNR/NBT-NS poisoner for credential capture \\
    & \texttt{evil-winrm} & WinRM shell with pass-the-hash support \\
    & \texttt{chisel} & HTTP tunneling for network pivoting \\
    & \texttt{proxychains} & SOCKS/HTTP proxy routing for pivoting \\
\midrule
\multirow{5}{*}{Credential Attacks}
    & \texttt{hashcat} & GPU password cracker (300+ hash types) \\
    & \texttt{john} & Rule-based password cracker \\
    & \texttt{hydra} & Online bruteforcing (50+ protocols) \\
    & \texttt{impacket} & Protocol library (secretsdump, psexec, wmiexec) \\
    & \texttt{kerbrute} & Kerberos user enumeration and password spraying \\
\midrule
\multirow{8}{*}{Active Directory}
    & \texttt{bloodhound} & AD attack path visualization via graph analysis \\
    & \texttt{sharphound} & BloodHound data collector \\
    & \texttt{rubeus} & Kerberos attack toolkit (roasting, tickets) \\
    & \texttt{mimikatz} & Memory credential extraction \\
    & \texttt{powerview} & AD enumeration PowerShell tool \\
    & \texttt{ldapdomaindump} & LDAP data extraction \\
    & \texttt{pingcastle} & AD security assessment and risk scoring \\
    & \texttt{adrecon} & AD reconnaissance reporting \\
\midrule
\multirow{4}{*}{Privilege Escalation}
    & \texttt{linpeas} & Linux privesc enumeration \\
    & \texttt{winpeas} & Windows privesc enumeration \\
    & \texttt{pspy} & Linux process monitor (cron, scheduled tasks) \\
    & \texttt{seatbelt} & Windows security auditing \\
\bottomrule
\end{tabular}
\end{table*}

\section{Evidence Confidence Scoring}
\label{sec:appendix-evidence}

Table~\ref{tab:evidence-scoring} presents the complete evidence confidence scoring rubric used by the TDA mechanism.
Scores are assigned deterministically based on evidence type, enabling reproducible difficulty assessment.

\begin{table*}[h]
\centering
\caption{Evidence confidence scoring rubric. Scores are assigned based on the strongest evidence type at each node; when multiple evidence types are present, the highest applicable score is used.}
\label{tab:evidence-scoring}
\small
\begin{tabular}{lcp{8cm}}
\toprule
\textbf{Evidence Type} & \textbf{Score} & \textbf{Indicators} \\
\midrule
\multicolumn{3}{l}{\textit{Verified Evidence (Exploitation Confirmed)}} \\
Valid credentials & 1.0 & Successful authentication via SSH, WinRM, SMB, or web login \\
Shell access & 1.0 & Interactive command execution confirmed \\
Data exfiltration & 1.0 & Sensitive data retrieved (flags, database contents, config files) \\
\midrule
\multicolumn{3}{l}{\textit{Confirmed Vulnerability (Exploit Available)}} \\
CVE with public exploit & 0.8 & Vulnerability scanner confirmation + Exploit-DB/Metasploit module exists \\
Auth bypass confirmed & 0.8 & Endpoint accessible without credentials when authentication expected \\
Injection confirmed & 0.8 & SQL/command injection produces observable side effects \\
\midrule
\multicolumn{3}{l}{\textit{Plausible Hypothesis (Evidence Supports)}} \\
Version-matched vuln & 0.5 & Service version matches known vulnerable version range \\
Configuration weakness & 0.5 & Misconfiguration identified (default credentials, open permissions) \\
Information disclosure & 0.5 & Sensitive information leaked (usernames, paths, internal IPs) \\
\midrule
\multicolumn{3}{l}{\textit{Speculative Hypothesis (Minimal Evidence)}} \\
Service identified & 0.3 & Port open with service fingerprint, no version/vulnerability match \\
Potential attack surface & 0.3 & Endpoint exists but no vulnerability indicators \\
Unconfirmed assumption & 0.3 & Hypothesis based on common patterns without direct evidence \\
\bottomrule
\end{tabular}
\end{table*}

\noindent\textbf{Path Confidence Computation.}
For a path $P = (n_0, n_1, \ldots, n_k)$ from root to current node, the evidence confidence is computed as:
\begin{equation}
E(P) = \frac{1}{k} \sum_{i=1}^{k} e(n_i)
\end{equation}
where $e(n_i)$ is the confidence score assigned to node $n_i$ based on Table~\ref{tab:evidence-scoring}.
The root node $n_0$ is excluded as it represents the initial state before any evidence is gathered.

\noindent\textbf{Tool Output Parsing.}
Evidence types are determined automatically by parsing tool outputs against expected patterns.
For example, \texttt{nmap} output containing ``open'' with a service version triggers version-matched vulnerability lookup (0.5); \texttt{sqlmap} output containing ``injectable'' triggers confirmed injection (0.8); successful \texttt{ssh} connection triggers valid credentials (1.0).
The Tool Layer's typed interfaces (Section~\ref{sec:tool-layer}) provide structured outputs that simplify this parsing.

\noindent\textbf{Example.}
Consider a path: \emph{port scan} $\rightarrow$ \emph{web server (nginx 1.18)} $\rightarrow$ \emph{directory bruteforce} $\rightarrow$ \emph{login form discovered} $\rightarrow$ \emph{SQL injection confirmed}.
Evidence scores are: 0.3 (service identified), 0.5 (version-matched to known nginx vulnerabilities), 0.3 (endpoint exists), 0.8 (injection confirmed).
Path confidence $E = (0.3 + 0.5 + 0.3 + 0.8)/4 = 0.475$, indicating moderate confidence appropriate for transitioning from reconnaissance to exploitation.

\section{Parameter Derivation and Validation}
\label{sec:appendix-parameters}

This appendix documents the derivation and sensitivity analysis for hyperparameters in \tool.

\subsection{Validation Dataset}

All hyperparameters are tuned on a held-out validation set of 30 execution traces from retired HTB machines (2022--2023), disjoint from the PentestGPT Benchmark evaluation set.
The validation set includes 10 Easy, 12 Medium, and 8 Hard machines, covering web exploitation (12), Linux privilege escalation (10), and Windows/AD attacks (8).
We use GPT-4o for validation to avoid overlap with evaluation models (GPT-5.2, Opus 4.5, Gemini 3).

\begin{table}[h]
\centering
\caption{TDI weight sensitivity analysis. Performance (subtask completion \%) across weight configurations. Bold indicates selected weights.}
\label{tab:weight-sensitivity}
\small
\begin{tabular}{cccc|c}
\toprule
$w_H$ & $w_E$ & $w_C$ & $w_S$ & Performance (\%) \\
\midrule
0.25 & 0.25 & 0.25 & 0.25 & 71.2 \\
\textbf{0.30} & \textbf{0.30} & \textbf{0.20} & \textbf{0.20} & \textbf{73.8} \\
0.35 & 0.25 & 0.20 & 0.20 & 72.4 \\
0.25 & 0.35 & 0.20 & 0.20 & 73.1 \\
0.30 & 0.25 & 0.25 & 0.20 & 72.9 \\
0.40 & 0.30 & 0.15 & 0.15 & 70.8 \\
\bottomrule
\end{tabular}
\end{table}

\subsection{TDI Weight Selection}

Table~\ref{tab:weight-sensitivity} presents TDI weights derived via grid search over $w \in [0.1, 0.4]$ with step size 0.05, subject to $\sum w_i = 1$.
Performance is measured as mean subtask completion rate across the validation set.

Performance varies within $\pm$3\% across configurations where all weights remain in $[0.1, 0.4]$, indicating robustness to precise weight selection.
The selected configuration ($w_H = w_E = 0.3$, $w_C = w_S = 0.2$) reflects domain intuition: horizon and evidence confidence are primary difficulty signals, while context load and success rate provide secondary modulation.

\begin{table}[h]
\centering
\caption{Mode selection threshold sensitivity. Performance (subtask completion \%) across threshold configurations.}
\label{tab:threshold-sensitivity}
\small
\begin{tabular}{cc|c}
\toprule
$\theta_{\text{explore}}$ & $\theta_{\text{exploit}}$ & Performance (\%) \\
\midrule
0.5 & 0.2 & 72.1 \\
0.5 & 0.3 & 72.8 \\
0.6 & 0.2 & 73.2 \\
\textbf{0.6} & \textbf{0.3} & \textbf{73.8} \\
0.6 & 0.4 & 72.4 \\
0.7 & 0.3 & 73.0 \\
0.7 & 0.4 & 71.6 \\
\bottomrule
\end{tabular}
\end{table}

\subsection{Mode Selection Thresholds}

Table~\ref{tab:threshold-sensitivity} presents sensitivity analysis for mode selection thresholds ($\theta_{\text{explore}}$, $\theta_{\text{exploit}}$).

The intermediate zone ($\theta_{\text{exploit}} \leq \text{TDI} \leq \theta_{\text{explore}}$) triggers \textsc{LLMDecide}.
Narrower zones reduce LLM calls but sacrifice adaptivity; wider zones increase overhead without proportional benefit.

\subsection{Pruning Parameters}

The pruning threshold ($\theta_{\text{prune}} = 0.8$) and minimum attempts ($k_{\min} = 3$) prevent both premature and excessively delayed pruning.

\begin{table}[h]
\centering
\caption{Pruning parameter sensitivity. Metrics: subtask completion (\%), branches incorrectly pruned (\%), wasted attempts on intractable branches (mean count).}
\label{tab:prune-sensitivity}
\small
\begin{tabular}{cc|ccc}
\toprule
$\theta_{\text{prune}}$ & $k_{\min}$ & Completion & False Prune & Wasted \\
\midrule
0.7 & 2 & 71.2 & 8.4 & 2.1 \\
0.7 & 3 & 72.4 & 5.2 & 3.4 \\
\textbf{0.8} & \textbf{3} & \textbf{73.8} & \textbf{2.8} & \textbf{4.1} \\
0.8 & 4 & 73.2 & 1.9 & 5.8 \\
0.9 & 3 & 72.1 & 1.2 & 6.9 \\
\bottomrule
\end{tabular}
\end{table}

Lower thresholds increase false pruning (abandoning tractable paths); higher thresholds waste attempts on intractable paths.
The selected configuration achieves favorable balance.

\subsection{UCB Difficulty Penalty}

The difficulty penalty coefficient ($\lambda = 0.5$) modulates how strongly TDI affects node selection in the UCB formula.

\begin{table}[h]
\centering
\caption{UCB difficulty penalty ($\lambda$) sensitivity.}
\label{tab:lambda-sensitivity}
\small
\begin{tabular}{c|cc}
\toprule
$\lambda$ & Completion (\%) & Backtrack Rate (\%) \\
\midrule
0.0 (standard UCB) & 68.4 & 12 \\
0.25 & 71.2 & 21 \\
\textbf{0.5} & \textbf{73.8} & \textbf{34} \\
0.75 & 72.1 & 42 \\
1.0 & 69.8 & 51 \\
\bottomrule
\end{tabular}
\end{table}

$\lambda = 0$ recovers standard UCB, which underperforms due to insufficient difficulty awareness.
$\lambda = 1.0$ over-penalizes difficult nodes, preventing exploration of challenging but tractable paths.

\subsection{Context Load Degradation Study}
\label{sec:appendix-context}

To establish the 40\% context load threshold, we conduct a controlled study measuring LLM instruction-following accuracy under varying context loads.

\noindent\textbf{Methodology.}
We construct 50 penetration testing instruction-following tasks from an independent GOAD deployment (separate from evaluation instances).
Each task comprises a system state description, accumulated context (tool outputs, discovered information), and a specific instruction (e.g., ``Extract the service account password from the Kerberoast output and attempt authentication'').
Tasks are designed with unambiguous correct responses, enabling binary accuracy scoring.

For each task, we generate context variants at 10\%, 20\%, 30\%, 40\%, 50\%, 60\%, 70\%, 80\%, and 90\% of the model's context window.
Context padding uses realistic penetration testing artifacts: verbose tool outputs, reconnaissance results, and session histories from actual GOAD runs.
Padding is inserted before the instruction to simulate accumulated session context.
We evaluate GPT-4o (128K context), Claude-3-Sonnet (200K context), and Gemini-1.5-Pro (1M context) with temperature 0, running each task-context combination three times.

Performance remains stable ($>$90\%) up to 40\% load, then degrades approximately linearly.
The 40\% threshold represents the inflection point beyond which additional context yields diminishing returns and begins actively harming performance.

\noindent\textbf{Failure Mode Analysis.}
Beyond 40\% load, failures concentrate in three categories: ignoring relevant information from earlier context (42\% of failures), hallucinating tool outputs not present in context (31\%), and executing incorrect but plausible commands (27\%).
These patterns align with the ``lost in the middle'' phenomenon documented in prior work~\cite{liu2024lost}.




\end{document}